\documentclass[
 preprint,
 superscriptaddress,
 amsmath,amssymb,
 aps,
 prmaterials,
 showpacs
]{revtex4-1}

\usepackage{graphicx}% Include figure files
\usepackage{dcolumn}% Align table columns on decimal point
\usepackage{bm}% bold math

\begin{document}

%\preprint{APS/123-QED}

\title{Many-Body Effect of Mesoscopic Localized States in MoS$_2$ Monolayer}

\author{Yang Yu}
\thanks{Contributed equally to this work.}
\author{Jianchen Dang}
\thanks{Contributed equally to this work.}
\author{Chenjiang Qian}
\thanks{Contributed equally to this work.}
\author{Sibai Sun}
\author{Kai Peng}
\author{Xin Xie}
\author{Shiyao Wu}
\author{Feilong Song}
\author{Jingnan Yang}
\author{Shan Xiao}
\author{Longlong Yang}
\affiliation{Beijing National Laboratory for Condensed Matter Physics, Institute of Physics, Chinese Academy of Sciences, Beijing 100190, China}
\affiliation{CAS Center for Excellence in Topological Quantum Computation and School of Physical Sciences, University of Chinese Academy of Sciences, Beijing 100049, China}
\author{Yunuan Wang}
\affiliation{Beijing National Laboratory for Condensed Matter Physics, Institute of Physics, Chinese Academy of Sciences, Beijing 100190, China}
\affiliation{Key Laboratory of Luminescence and Optical Information, Ministry of Education, Beijing Jiaotong University, Beijing 100044, China}
\author{Xinyan Shan}
\affiliation{Beijing National Laboratory for Condensed Matter Physics, Institute of Physics, Chinese Academy of Sciences, Beijing 100190, China}
\author{M. A. Rafiq}
\affiliation{Department of Physics and Applied Mathematics, Pakistan Institute of Engineering and Applied Sciences, P.O. Nilore Islambad Pakistan}
\author{Bei-Bei Li}
\affiliation{Beijing National Laboratory for Condensed Matter Physics, Institute of Physics, Chinese Academy of Sciences, Beijing 100190, China}
\author{Xiulai Xu}
\email{xlxu@iphy.ac.cn}
\affiliation{Beijing National Laboratory for Condensed Matter Physics, Institute of Physics, Chinese Academy of Sciences, Beijing 100190, China}
\affiliation{CAS Center for Excellence in Topological Quantum Computation and School of Physical Sciences, University of Chinese Academy of Sciences, Beijing 100049, China}
\affiliation{Songshan Lake Materials Laboratory, Dongguan, Guangdong 523808, China}

\date{\today}

\begin{abstract}

Transition metal dichalcogenide monolayers provide an emerging material system to implement quantum photonics with intrinsic two-dimensional excitons or embedded zero-dimensional localized states. Here we demonstrate the mesoscopic localized states between two- and zero- dimensions, which is a many-body system with electron-electron Coulomb interactions. A fine structure splitting is observed, which is similar to quantum dots. Meanwhile the polarization is changed by the magnetic field, due to the nature of two-dimensional monolayers. Furthermore, a large quadratic diamagnetism with a coefficient of around $100\ \mathrm{\mu eV/T^2}$  is observed, as a unique consequence of the mesoscopic scale. The many-body effect also results in the emission energy variation and linewidth narrowing in the spectrum, which corresponds well to the theoretical analysis. These unique properties indicate the great potential of mesoscopic localized states in many-body physics and quantum photonics.

\end{abstract}

\maketitle

%\section{\label{sec1}Introduction \protect\\}

Monolayer transition metal dichalcogenides (TMDCs) as direct-gap semiconductors, such as MoS$_2$ and WS$_2$ with a band gap in the visible region, offer a practical platform for quantum photonics due to their distinctive electronic and optical properties \cite{PhysRevLett.108.196802,PhysRevLett.105.136805,doi:10.1021/nl903868w}. In imperfect monolayers, the defect-induced localized states usually have characteristics related to the spatial extent. Zero-dimensional localized states with small spatial extent are similar to semiconductor quantum dots (QDs) \cite{Trauzettel2007,Ponomarenko356,PhysRevLett.105.116801}. While as the spatial extent becomes larger, the inherent two-dimensional material properties dominate in the localized state. Therefore, the mesoscopic localized state between two- and zero- dimensions can be predicted, which might have characteristics from both systems and additional unique optical properties.

In contrast to two-dimensional degenerate Landau levels and zero-dimensional discrete Fock-Darwin levels \cite{PhysRevLett.98.186803,PhysRevB.77.235411}, energy levels of mesoscopic localized states are discrete and dense. The multiple electrons occupied in the levels with electron-electron Coulomb interactions result in a many-body system \cite{Dzyubenko2003,Kleemans2010,PhysRevLett.106.046401}. Although the many-body effect is complex \cite{Almand-Hunter2014,PhysRevMaterials.2.124003}, interesting properties can be predicted, such as anticrossings between energy levels and variations in the spectra \cite{PhysRevB.93.165410,PhysRevB.96.115428}.

Here, we experimentally demonstrate the mesoscopic localized states in the MoS$_2$ monolayer. A wide ensemble of photoluminescence (PL) peak with energy smaller than excitons and trions demonstrates the defect-induced localized state \cite{Mak2012,PhysRevB.88.045318,doi:10.1063/1.4954837}. An energy splitting similar to the fine structure splitting (FSS) in QDs is observed \cite{RevModPhys.85.79}, originating from the asymmetric defect interface. When the magnetic field is applied, the polarization is significantly changed, as the feature of $K$ and $K'$ valleys located at the two-dimensional hexagonal Brillouin zone \cite{Barbone2018,PhysRevB.77.235406}. Furthermore, the PL spectra show a large quadratic diamagnetism, as a consequence of the many-body system with a mesoscopic scale. The Coulomb interactions in the many-body system concurrently result in the level anticrossings and linewidth narrowing with the increasing magnetic field. Our work demonstrates the localized states in monolayer MoS$_2$ with both two- and zero- dimensional features along with additional unique optical properties. Such mesoscopic system expands the investigations on TMDCs monolayers for many-body physics and has great potential in future applications in quantum optoelectronics.

%\section{\label{sec2}Mesoscopic States \protect\\}

\begin{figure}
\centering
\includegraphics[scale=0.8]{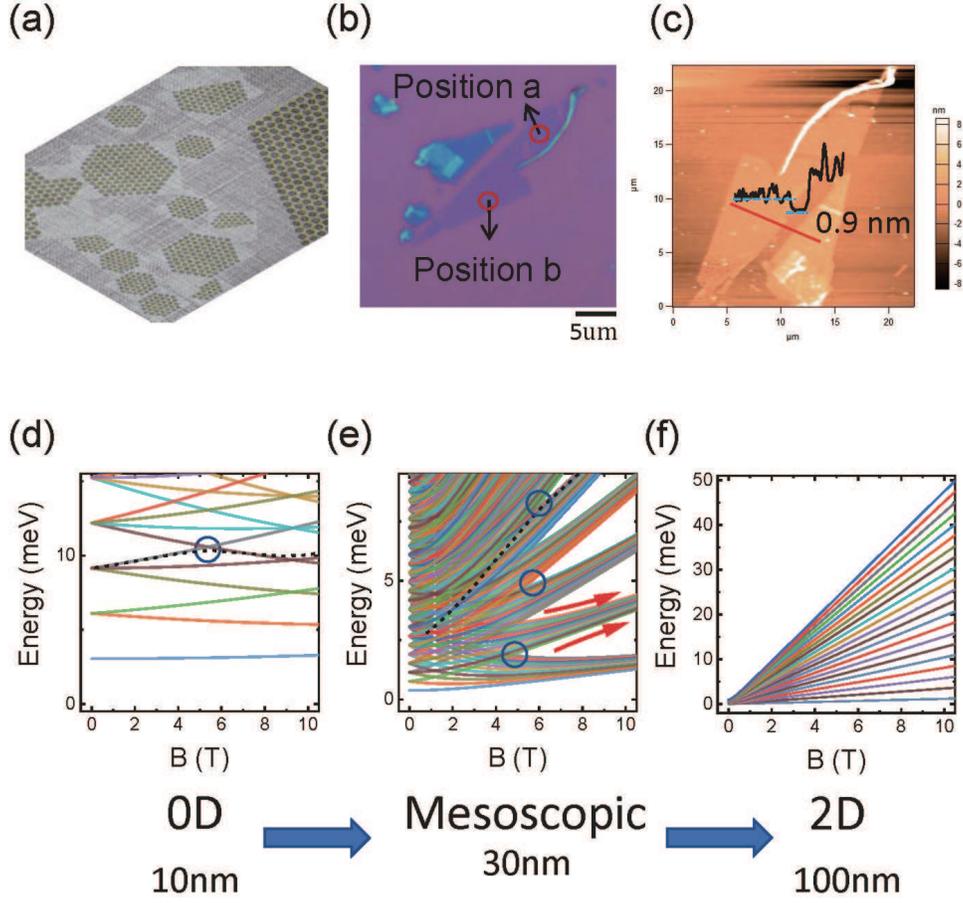}
\caption{(a) Schematic diagram of localized states with different scales. (b) The optical microscopy image of the sample. (c) The AFM image of the sample.The thinkness profile line is shown in inset and the scale bars are 5 $\mu$m. (d)-(f) Calculated Fock-Darwin energy levels with different particle sizes. (d) Size of 10 nm with discrete levels. (e) Size of 30 nm as mesoscopic states between zero- and two- dimensions. (f) Size of 100 nm with typical two-dimensional Landau levels. Black dashed lines and blue circles refer to level crossings due to Coulomb interactions. Red arrows refer to the energy band width shrinking with the magnetic field.}
\label{p1}
\end{figure}

The monolayer of MoS$_2$ as a two-dimensional material could contain various defect-induced localized states which could originate from bilayer parts, atom losses or monolayer flakes \cite{Tongay2013,doi:10.1021/nn5073495}.  A schematic diagram of localized states with different sizes is shown in Fig.~\ref{p1}(a). The optical microscopy and atomic force microscope (AFM) images of our sample are shown in Fig.~\ref{p1}(b) and (c). Defects with different spatial extent can be resolved with the AFM profile line. Previous works mainly focused on the intrinsic monolayers \cite{Koperski_2018,Mak:2012aa,PhysRevLett.111.216805} or the small defects as zero-dimensional quantum emitters \cite{Srivastava2015,Koperski2015,Chakraborty2015,He2015}. Normally, the Fock-Darwin levels of the ideal two-dimensional harmonic oscillator are a good model to qualitatively describe the features of these systems with a magnetic field \cite{PhysRevB.77.235411,PhysRevX.4.011034}. To briefly show the localized states tending from zero- to two- dimension, we calculate the Fock-Darwin levels with the size $l_{xy}$ of 10, 30 and 100 nm as shown in Fig.~\ref{p1}(d)-(f). The effective mass is set with $0.5\ m_e$ as a typical value for MoS$_2$ monolayers \cite{PhysRevB.85.033305,ShengYu_effectivemass}. The results with $l_{xy}=10\ \mathrm{nm}$ (Fig.~\ref{p1}(d)) show the features of zero-dimensional system similar to QDs. There are many level crossings as marked by the blue circle. Due to the electron-electron Coulomb interactions as the many-body effect, these level crossings would result in anticrossings (black dashed line in Fig.~\ref{p1}(d)) in the spectra \cite{PhysRevB.93.165410}. For $l_{xy}=100\ \mathrm{nm}$ (Fig.~\ref{p1}(f)) the linear and degenerate Landau levels can be observed, as a typical feature of two-dimensional materials. While for a mesoscopic scale between zero- and two- dimension with $l_{xy}=30\ \mathrm{nm}$ (Fig.~\ref{p1}(e)), the results are much more complex with discrete and dense energy levels. As the magnetic field becomes larger, the system shows degenerate Landau levels, resulting in many level crossings marked by blue circles. Similarly, due to the many-body effect, the inflection of the emission energy (black dashed line in Fig.~\ref{p1}(e)) could be predicted, indicating the interaction between an inherent two-dimensional system and a finite defect system \cite{Dzyubenko2003,PhysRevLett.103.046810}. Meanwhile, as the magnetic field increases, the discrete energy levels contract to degenerate Landau levels (marked by red arrows). This could be considered as the shrinking of the energy `bands', an unique feature of the mesoscopic localized states.

%\section{\label{sec3}Results and Discussion \protect\\}

The optical spectroscopy measurement was carried out in a cryogenic vacuum system. The sample was cooled down to 4.2 K by liquid helium and excited by a laser with a wavelength of 532 nm. A magnetic field was applied by the superconducting magnets. The PL spectra were collected at several positions on the sample by a spectrometer with the resolution of 0.1 nm. More detailed information about sample preparation and measurement is shown in the Supplementary Information \cite{sl}.

\begin{figure}
\centering
\includegraphics[scale=0.8]{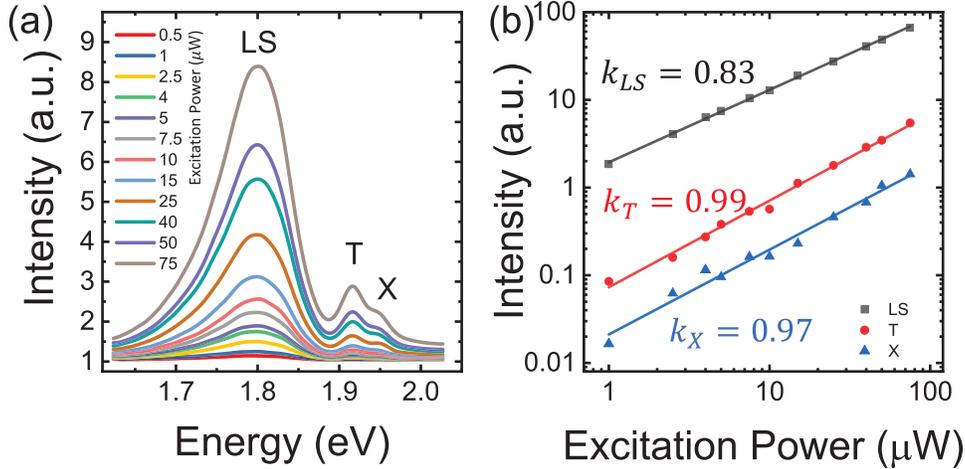}
\caption{(a) Power-dependent PL spectra of the MoS$_2$ monolayer. (b) The logarithmic dependence of peak intensities on excitation power. $k_{LS}$ is 0.8, between typical values of 1 for two-dimensional materials and 0.6 for zero-dimensional localized states.}
\label{p2}
\end{figure}

Figure \ref{p2}(a) shows the power-dependent PL spectra of one position (position a (shown in Fig.~1(b))) on the sample. The excitation power was changed from 0.5 $\mu$W to 75 $\mu$W. Three peaks can be clearly distinguished in the spectra, with the emission energies around 1.80, 1.92 and 1.95 eV. Compared with previous works, it can be concluded that three peaks correspond to the recombinations from the localized states (LS), trion (T) and exciton (X) states  respectively \cite{Mak2012,PhysRevB.88.045318,doi:10.1063/1.4954837}. Figure \ref{p2}(b) shows the logarithmic linear dependence of PL intensities on the excitation power. $k_T$ and $k_X$ are both around 1, which is typical excitonic states in the two-dimensional monolayers \cite{doi:10.1021/nn5059908}. However, $k_{LS}$ is 0.83, different from either the exciton (value of 1) or zero-dimensional localized states (typical value of 0.6) \cite{doi:10.1021/nn5059908}. Meanwhile, the linewidth of peak LS is around 0.12 eV, much broader than that of zero-dimensional localized states in TMDCs monolayers \cite{Srivastava2015,Koperski2015,He2015,Chakraborty2015}. The slope and linewidth indicate an unconventional defect-induced localized state, which could be introduced by defects in a mesoscopic scale.

 %This between two- and zero- dimensional mesoscopic localized state should have the interesting optical properties as discussed above theoretically.

\begin{figure}
\centering
\includegraphics[scale=0.8]{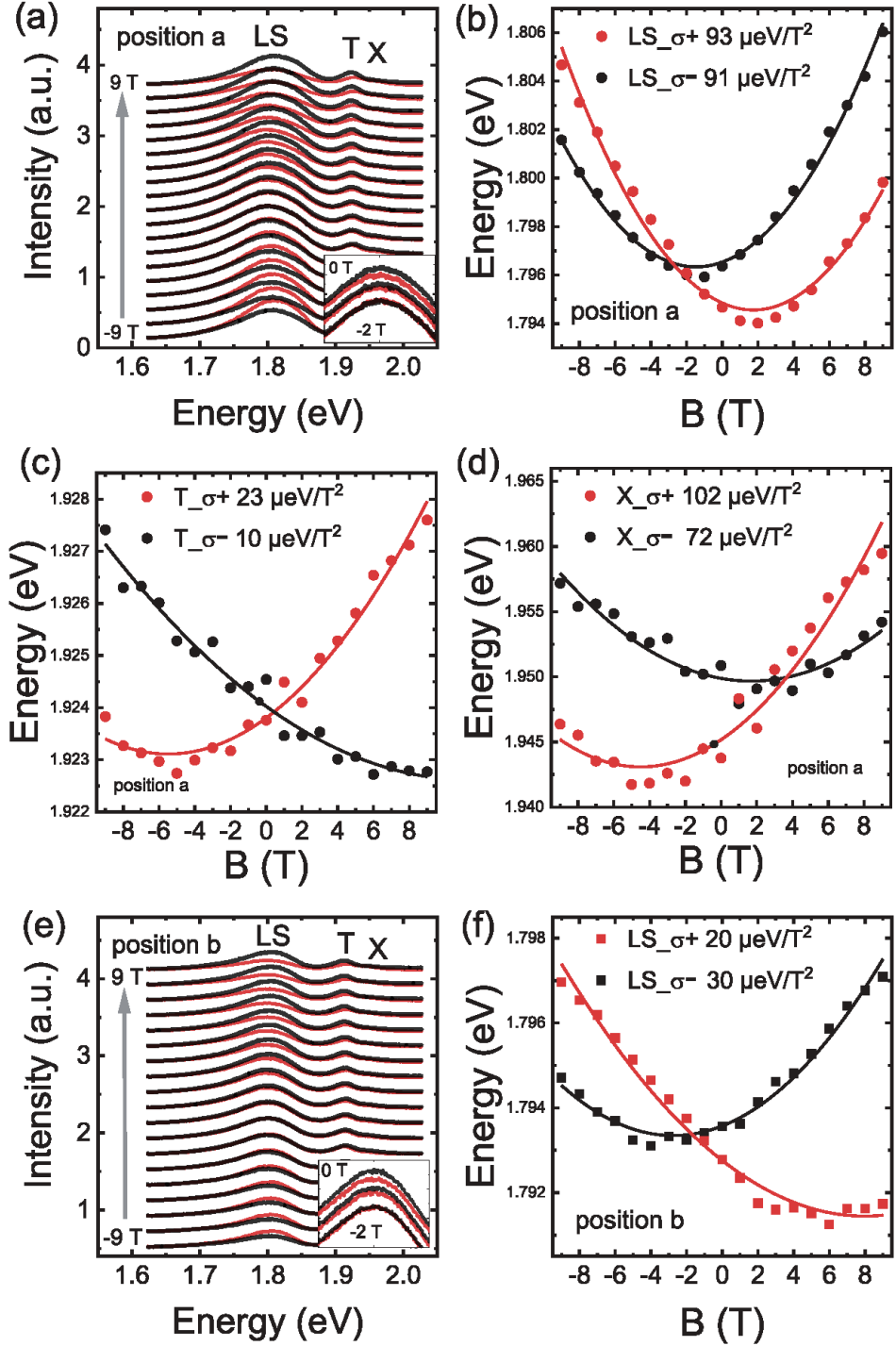}
\caption{(a) PL spectra at position a with $\sigma^-$ and $\sigma^+$ polarized detection under linearly polarized excitation. (b)-(d) The diamagnetism of L (b), T (c) and X (d) in PL spectra at position a. The quadratic fittings in (c) and (d) are not very well. (e) PL spectra measured at position b with $\sigma^-$ and $\sigma^+$ polarized detection. (f) The diamagnetism of L at position b. Insets in (a) and (e) clearly show the splitting at $B=0$.}
\label{p3}
\end{figure}

To further investigate the mesoscopic localized state, the magneto-PL spectra were measured under the excitation power of 75 $\mu$W. A vertical magnetic field (vertical to the monolayer surface) $B_\perp$ was changed from -9 to 9 T, and the polarized PL spectra were collected by using a quarter-wave plate and a linear polarizer. Measurement was carried out at position a (Fig.~\ref{p3}(a)) and repeated at different positions (Fig.~\ref{p3}(e)). The black lines refer to the quarter-wave plate at $-\pi/4$ ($\sigma^-$) corresponding to $K'$ valley emission, and red lines refer to the quarter-wave plate at $\pi/4$ ($\sigma^+$) before the linear polarizer, corresponding to $K$ valley emission \cite{Mak2012}. Interestingly, the emission energy is asymmetric with the sign of $B_\perp$, and an FSS between $\sigma^-$ and $\sigma^+$ at $B_\perp=0$ was observed. The FSS is usually observed in QDs due to the asymmetric structure \cite{RevModPhys.85.79}. Similarly, we ascribe these anomalous phenomena to the defect-induced layer symmetry breaking and the lift of valley-emission degeneracy, as a similar feature of zero-dimensional system \cite{PhysRevB.93.165410}. Meanwhile, the polarization of all three peaks changed dramatically with the magnetic field, due to the thermal equilibrium-related valley property \cite{Barbone2018}. This is the feature of two-dimensional materials and has not been observed in zero-dimensional localized states \cite{Tongay2013,Srivastava2015,Koperski2015,He2015,Chakraborty2015}.

Furthermore, $\sigma^-$ and $\sigma^+$ emissions both shift towards higher energy in the presence of $B_\perp$, exhibiting an analogous quadratic diamagnetism to conventional QDs \cite{PhysRevLett.101.267402,PhysRevB.81.113307,Cao2015,Cao2016} as shown in Fig.~\ref{p3}(b). More importantly, at position a, both $\sigma^-$ and $\sigma^+$ polarized localized state exhibit large quadratic diamagnetic coefficient of around $100\ \mathrm{\mu eV/T^2}$, an extremely large value compared with the results in TMDCs or QDs \cite{Stier2016}. The diamagnetic effects for trion and exciton do not reveal good field-dependent regularity as shown in Fig.~\ref{p3}(c) and (d). Theoretically, their energy shift should be linear to $B_\perp$ in defectless monolayers \cite{Stier2016,PhysRevLett.121.057402}. The difference in the experiment might originate from the different defect densities in the monolayer but the linear feature defectless monolayers also affect our results with asymmetric parabolic lines observed. The quadratic diamagnetism of localized states is repeatable at position b, as shown in Fig.~\ref{p3}(f). Different diamagnetic coefficients for positions a and b can be attributed to different defect-induced localized states, depending on the distribution, shape, edge and size of defects. The theoretical model of the quadratic diamagnetic coefficient as $e^2 l_\alpha^2/8 m_\alpha $ is widely used to explore the diamagnetic phenomenon, where $m_\alpha$ is the effective mass and $l_\alpha$ is the lateral extent of wave-function \cite{Cao2015}. The large diamagnetic coefficient of $100\ \mathrm{\mu eV/T^2}$ in Fig.~\ref{p3}(b) corresponds to the lateral extent of around 30 nm, which verifies our assumption of the recombination from the mesoscopic states shown in Fig.~\ref{p1}(e).
%Given the difference between the initial state consisting of one electron-hole pair and final state after recombination

\begin{figure}
\centering
\includegraphics[scale=0.8]{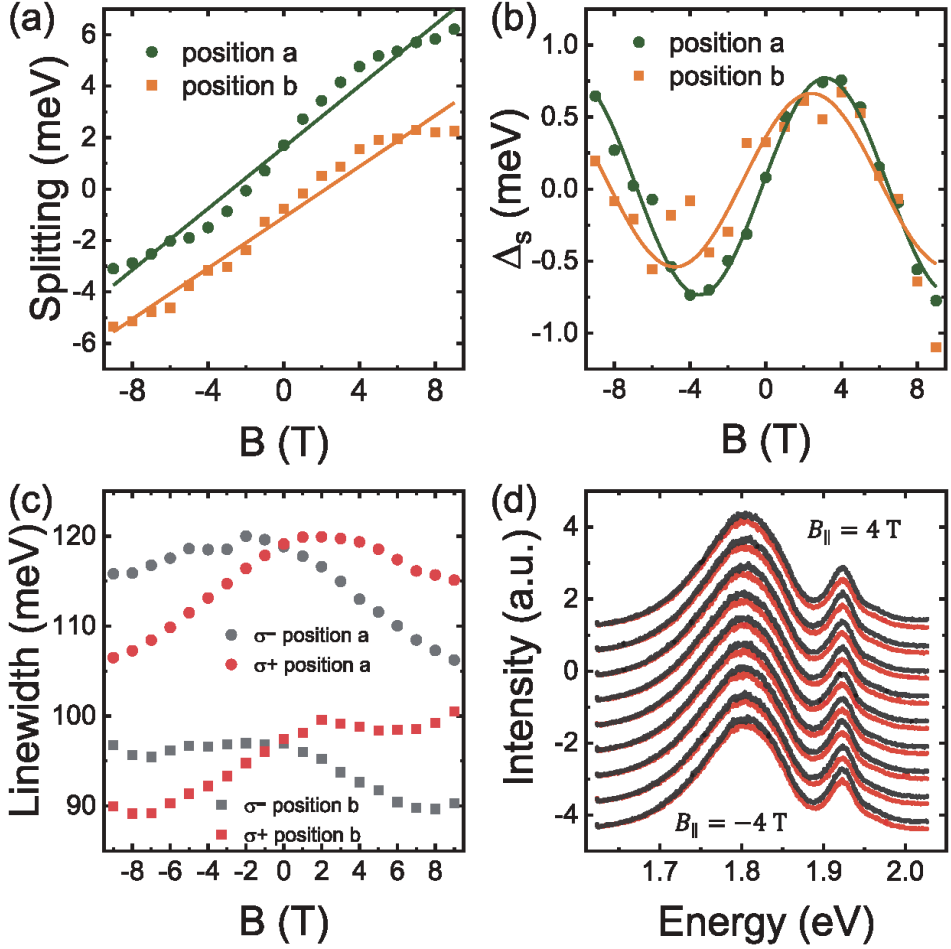}
\caption{(a) The energy splittings between $\sigma^-$ and $\sigma^+$ polarized localized states at two positions. The results generally indicate a linear g-factor.  (b) The variations between energy splittings and linear g-factor results in (a), mainly originating from the level crossings when the mesoscopic system shrinks into Landau levels as shown by the blue circles in Fig.~\ref{p1}(c)(middle).  (c) The linewidth narrowing with the magnetic field, which is also due to the mesoscopic scale as shown by red arrows in Fig.~\ref{p1}(e). (d) Polarized PL spectra with an in-plane magnetic field. No changes can be resolved, which is due to the little wave-function extent in the vertical direction for two-dimensional materials.}
\label{p4}
\end{figure}

The quadratic fitting in Fig.~\ref{p3}(b) and (f) have differences to the experimental data. To clearly show the phenomenon behind quadratic diamagnetism, we extracted the energy splitting between $\sigma^-$ and $\sigma^+$ emission at both positions as shown in Fig.~\ref{p4}(a). Generally, the Zeeman splittings show a linear dependence with the g-factor of around -9. This value is in accordance with previously reported localized states in TMDCs \cite{Srivastava2015,Koperski2015,Chakraborty2015}. However, obvious nonlinear terms can also be observed in Fig.~\ref{p4}(a), revealing the further phenomenon behind the quadratic diamagnetism and Zeeman splitting. The nonlinear terms, as the difference between experimental data and linear fittings, are shown in Fig.~\ref{p4}(b). Within the measurement range from -9 to 9 T, the nonlinear terms indicate a `periodic' variation with half period of around 7 T with small $B_\perp$. While at large $B_\perp$ around $9\ \mathrm{T}$, the nonlinear terms become deviated from the sine fitting. The variation of emission energy mainly originates at level crossings due to the electron-electron Coulomb interactions in the many-body system. As mentioned in the theoretical analysis above, there are two mechanisms for the level crossings in the localized states in TMDCs monolayer. One is the level crossing with interaction between localized electrons or holes as shown in the blue circle in Fig.~\ref{p1}(c). These level crossings are dense and usually with the period around 0.1 T \cite{PhysRevB.45.11419,PhysRevLett.71.613}. The other is the level crossing with the interaction between inherent two-dimensional system and finite defect structure as shown in blue circles in Fig.~\ref{p1}(d). These interactions usually occur at an interface around 0-8 T for different energy levels \cite{Dzyubenko2003,PhysRevLett.103.046810}. In zero-dimensional localized states, the two mechanisms are generally mixed and thus hard to distinguish as shown in previous work \cite{PhysRevLett.103.046810}. In contrast, the mesoscopic localized state with the broad linewidth is the ensemble of dense energy levels. Therefore, the dense level crossings arising from the first mechanism is averagely erased, and the level crossings from the second mechanism occur around 7 T as the average value of several levels. This non-linear variation demonstrates the many-body system with interaction between the mesoscopic localized states and the two-dimensional Landau levels. More detailed information on the non-linear variation of each valley and other fitting results can be seen in the Supplementary Information \cite{sl}.

Another unique optical property of the mesoscopic localized state with the magnetic field is the linewidth narrowing, as shown in Fig.~\ref{p4}(c) for both positions. The linewidths are around $50\ \mathrm{meV}$ for two-dimensional delocalized excitons and around $1\ \mathrm{meV}$ for zero-dimensional localized states, mainly originating from the finite lifetime of energy levels \cite{Srivastava2015,Koperski2015,Chakraborty2015,doi:10.1021/nn5059908}. As the non-radiative recombination increases, the linewidths usually broaden slowly with the magnetic field \cite{Stier2016,Barbone2018}. While for the mesoscopic localized state, the linewidth is large with values around $100\ \mathrm{meV}$. The broad linewidth originates from the `bandwidth' of ensemble of dense levels rather than the level lifetime as mentioned in the theoretical analysis above. Although intricate with the small $B_\perp$, the linewidth significantly narrows with the large $B_\perp$ when the system changes toward two-dimensional Landau levels as marked by red arrows in Fig.\ref{p1}(e), which can be observed in Fig.~\ref{p4}(c). Additionally, An in-plane magnetic field $B_\parallel$ up to 4 T has been applied to investigate the diamagnetism. It can be seen that the PL basically keeps invariable (Fig.~\ref{p4}(d)), which is due to the small wave-function extent in the vertical direction. This result agrees well with previous reports  \cite{Zhang:2017aa}, although the in-plane magnetic might brighten the dark exciton in Voigt geometry \cite{Molas_2017}.

%\section{\label{sec4}Conclusion \protect\\}

In summary, we have demonstrated the observation of mesoscopic localized states in MoS$_2$ monolayer with magnetophotoluminescence spectroscopy. With an applied magnetic field, the mesoscopic state has features from both zero- and two- dimensional systems. The polarization degree changes like a two-dimensional system, and the diamagnetism is quadratic like a zero-dimensional system. Furthermore, as a complex many-body system, the mesoscopic localized state has particular energy levels and optical properties. Large diamagnetism coefficients around $100\ \mathrm{\mu eV/T^2}$ have been obtained, which corresponds well to theoretical prediction. When the system changes towards the Landau levels, the nonlinear energy shift and the linewidth narrowing have been observed. The mesoscopic localized states in TMDs monolayers could provide a unique quantum photonic platform with a great potential in the future.

\begin{acknowledgments}

This work was supported by the National Natural Science Foundation of China under Grants No. 51761145104, No. 61675228, No. 11721404, and No. 11874419, the Strategic Priority Research Program (Grant No. XDB28000000), the Instrument Developing Project (Grant No. YJKYYQ20180036), the Interdisciplinary Innovation Team of the Chinese Academy of Sciences, and
the Key R$\&$D Program of Guangdong Province (Grant No. 2018B030329001).

\end{acknowledgments}

\end{document}